\documentclass[%
 reprint,
 amsmath,amssymb,
 aps,
]{revtex4-2}

\usepackage{graphicx}
\usepackage{dcolumn}
\usepackage{xcolor}
\usepackage{bm}

\newcommand{\n}{\boldsymbol{n}}
\newcommand{\no}{\boldsymbol{n_0}}
\newcommand{\strain}{\boldsymbol{\underline{\underline{\lambda}}}}

\newcommand{\xlink}{\boldsymbol{\underline{\underline{\ell}}}}
\newcommand{\lo}{\boldsymbol{\underline{\underline{\ell}}}_0}
\newcommand{\nn}{\nonumber}

\begin{document}
\title{Torsion and Bistability of Double-Twist Elastomers}

\author{Matthew P. Leighton}
\affiliation{Department of Physics, Simon Fraser University, Burnaby, British Columbia, V5A 1S6, Canada.}%
\author{Laurent Kreplak}
\author{Andrew D. Rutenberg}
\affiliation{Department of Physics and Atmospheric Science, Dalhousie University, Halifax, Nova Scotia, B3H 4R2, Canada.}%
\date{\today}

\begin{abstract}
We investigate the elastic properties of anisotropic elastomers with a  double-twist director field, which is a model for collagen fibrils or blue phases. We observe a significant Poynting-like effect, coupling torsion (fibril twist) and extension. For freely-rotating boundary conditions, we identify a structural bistability at very small extensional strains which undergoes a saddle-node bifurcation at a critical strain -- at approximately $1\%$ strain for a parameterization appropriate for collagen fibrils. With clamped boundary conditions appropriate for many experimental setups, the bifurcation is not present. We expect significant helical shape effects  when fixed torsion does not equal the equilibrium torsion of freely-rotating boundary conditions, due to residual torques.
\end{abstract}
\maketitle

\section{Introduction}
Covalently cross-linked elastomer networks are an important class of materials due to their extensibility, non-linear elasticity and biological and biomedical applications \cite{Hussain2021, Rohaley2022}. Liquid-crystalline elastomers are particularly interesting since their anisotropy is  tunable and affects other material properties \cite{Warner1996}. For example, liquid-crystalline elastomers can be used as actuators or sensors \cite{Ohm2010}. Of particular interest is how anisotropy affects the mechanical properties of liquid-crystal elastomers. 

Any liquid crystalline field can be incorporated into an anisotropic elastomer -- including the double-twist anisotropy of blue phases  \cite{tanaka2015double, schlafmann2021retention}. Collagen fibrils, the load bearing element of mammalian tissues, can be  modelled as an elastomeric cylinder of cross-linked collagen molecules arranged in such a double twist configuration~\cite{brown2014equilibrium, cameron2018polymorphism, Leighton2021non, Leighton2021chiral}. Keratin macrofibrils also exhibit a double-twist structure \cite{Harland2019}, and so should have similar elastomeric properties. 

Collagen fibrils can be mechanically tested both within their host tissue \cite{Misof1997, Bell2018} and \emph{ex vivo} \cite{Dutov2016, Quigley2018, Svensson2018, Yang2008, Liu2016}. Surprisingly, there is still no experimental consensus on the elastic properties of individual collagen fibrils for physiological strains below 10$\%$ \cite{Andriotis2022}. It is likely that varying levels of hydration, as controlled by osmotic pressure, explains some of the differences \cite{Andriotis2018}. Nevertheless, considerable variability is still observed within a single study \cite{Andriotis2022} -- indicating that fibrils may also differ in other respects. 

Intriguingly, collagen fibers (bundles of fibrils) rotate significantly under extensional strain \cite{Buchanan2017}. Such Poynting-like effects, coupling torsion and extension, are known to sensitively depend on material anisotropy \cite{Horgan2017, Horgan2016}. Helical shapes are also observed for collagen fibers \cite{Kalson2011, Thorpe2013, Buchanan2017}. Indeed, helical configurations (supercoiling) are observed for individual collagen fibrils \cite{Kalson2015}, as well as pairs \cite{Bozec2007, Safa2019} and larger groups \cite{Vidal2003, Franchi2010} of fibrils. Such supercoiling is often exhibited by cylindrical objects under torsion.

A recent theoretical study by Giudici and Biggins \cite{Giudici2021} showed that torsion effects in chiral cylinders can be significant, though in their study they focused on strain-torsion coupling with isotropic-nematic phase-transitions rather than mechanical effects of anisotropic elastomers. Their results emphasize that boundary conditions imposed on individual fibrils or collections of fibrils are important to consider. One source of the mechanical variation between fibrils may be differing torsional boundary conditions imposed by fibril assembly or during mechanical testing. 

Double-twist director fields require curvilinear coordinate systems and a theoretically-grounded deformation gradient tensor to dependably study elastic effects.  Fortunately, a differential-geometry approach for evaluating deformation gradient tensors within curvilinear coordinates exists \cite{melnikov2018deformation, Ogden1997}. With this, we can evaluate the deformation gradient tensor of a double-twist elastomer cylinder stretched along its axis while allowed to rotate.

In the simpler case where the collagen fibril is not allowed to rotate, we have already reported interesting results under extension or compression of double-twist elastomeric cylinders. Stretched collagen fibrils exhibit  director-field strain straightening \cite{bell2018hierarchical} that are captured by our double twist elastomer cylinder model \cite{Leighton2021d}. Collagen fibrils under compressive strain along their axis that show the sequential appearance of swollen domains along the fibrils~\cite{peacock2020buckling} can be understood in term of a phase coexistence between a high compression and a low compression state using the same model~\cite{Leighton2021chiral}. 

Nevertheless, not allowing fibril rotation (torsion) amounts to an uncontrolled assumption.  It is important to understand three implications of boundary conditions in this respect. First, natural  biological boundary conditions may not be clamped against rotation for the life of a fibril. Free rotation effects are therefore important to understand as a possible origin of observed developmental phenomena. Second,  experimental boundary conditions may vary between \emph{ex vivo} studies. Finally, if fibril ends are not allowed to rotate freely we anticipate that the resulting torque would couple with fibril shape to lead to visible supercoiling -- i.e. helical distortions of the fibril shape \cite{Giudici2021}.

To understand the potential roles of fibril torsion on the mechanical properties of collagen fibrils, or more generally of double-twist anisotropic elastomeric cylinders, we will study freely rotating (unclamped) fibrils under strain. This will determine the equilibrium fibril torsion vs strain. We also investigate clamped boundary conditions, where a specific non-zero torsion is imposed. Since we use a general elastomer free-energy, we can also investigate thermodynamically metastable mechanical equilibria related to torsion. Finally, we will describe how our results may manifest themselves in both \emph{in vivo} and \emph{ex vivo} studies. 

\section{Elastomeric Double-twist Model}
Consider a liquid-crystal elastomer with a molecular director field $\no$ and anisotropic Gaussian cross-links with density $\rho$. When this system is subject to a deformation, the entropic free energy density due to the cross-links is \cite{Warner1996}
\begin{equation}\label{anisotropic_fe}
f = \frac{\mu}{2} \textnormal{ Tr} (\lo \strain^\top \xlink^{-1} \strain).
\end{equation}
Here $\lo$ and $\xlink$ are tensors describing the initial and post-deformation orientations of the molecules and cross-links,  $\strain$ is the deformation-gradient tensor describing the deformation, $\mu=\rho k_B T$ is the shear modulus ($\rho$ is the molecular density, $k_B$ is Boltzmann's constant, and $T$ is the temperature). The tensors $\lo$ and $\xlink$ are functions of the pre- and post-strain molecular twist fields, and thus implicitly dependent on position; $\strain$ may also be position-dependent.

\begin{figure}[th]
\includegraphics[width=\columnwidth]{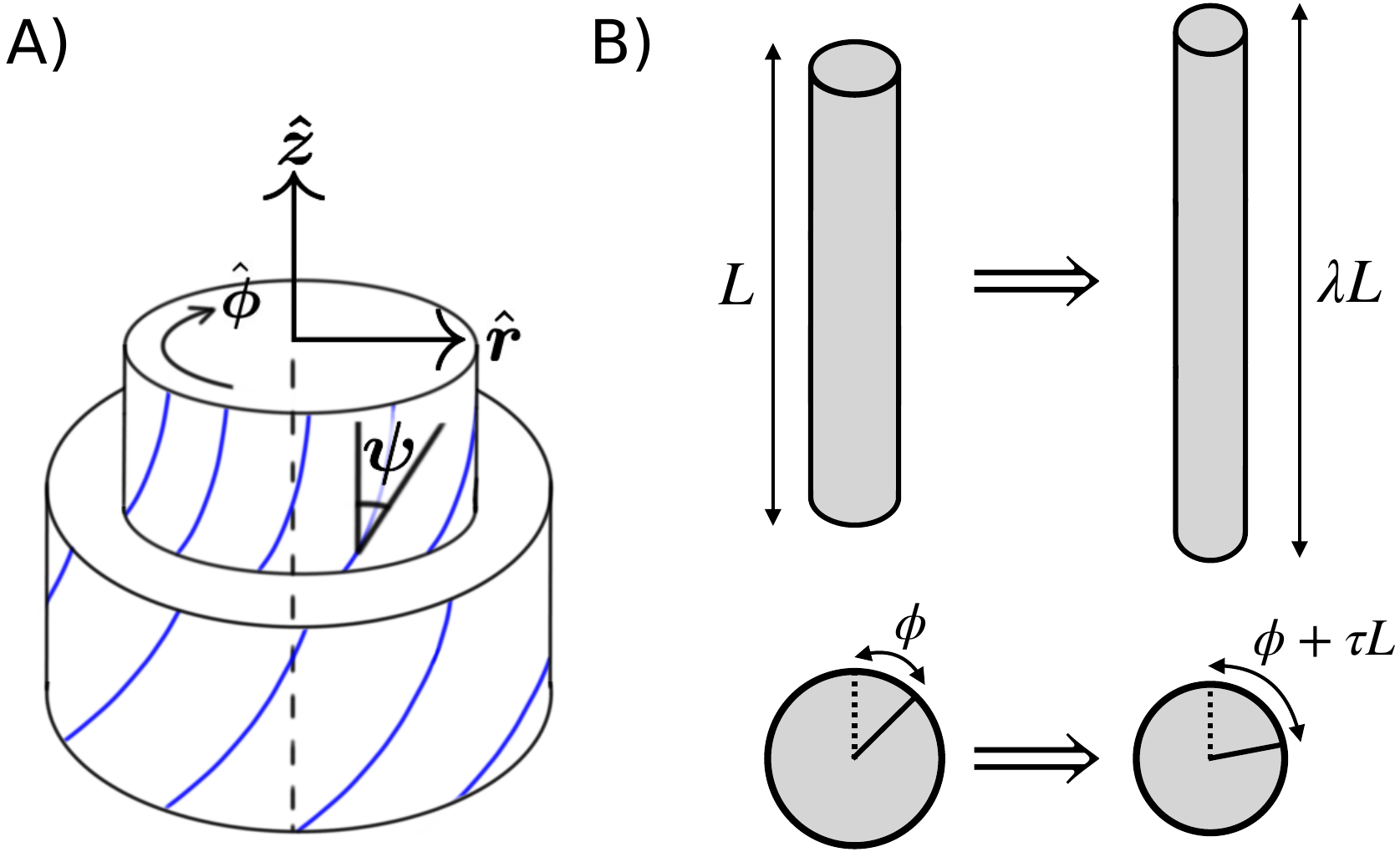}
\caption{\label{fig:diagram} A) Cutaway view of a double-twist cylinder, with the molecular director field indicated by curved blue lines. The three orthogonal directions in cylindrical coordinates, $\hat{\boldsymbol{r}}$, $\hat{\boldsymbol{\phi}}$, and $\hat{\boldsymbol{z}}$ are indicated. The twist angle $\psi(r)$, which is the angle of molecular tilt with respect to the axis, depends on the radial distance within the cylinder -- but is independent of $\phi$ or $z$. Adapted from \cite{Leighton2021chiral}. B) Schematic illustrating the axial deformation we consider. Seen from the side (above), the length is extended by a factor of $\lambda$ while the radius narrows due to incompressibility; (below) seen from the top, one end of the cylinder rotates by an angle $\tau L$ relative to the other end.}
\end{figure}

We are particularly interested in cylinders of unstrained anisotropic elastomers with a double-twist director field
\begin{equation}\label{initial_director}
\no(r) = -\sin\psi_0(r) \boldsymbol{\hat{\phi}} + \cos\psi_0(r) \boldsymbol{\hat{z}},
\end{equation}
which is found in collagen fibrils \cite{Leighton2021d, Leighton2021chiral} or keratin macrofibrils \cite{Harland2019} (Figure~\ref{fig:diagram}). The radius-dependent $\psi_0(r)$ is the initial twist angle function, which describes the angle of molecules with respect to the cylinder axis ($\boldsymbol{\hat{z}}$) before any mechanical strain is imposed. We assume azimuthal and translational symmetry, i.e. that the molecular tilt $\psi$ with respect to the cylinder axis does not depend on the azimuthal angle $\phi$ or on the axial coordinate $z$.  

For a cylindrical geometry, we will use the dimensionless radius $r=\tilde{r}/\tilde{R}$ in terms of the dimensioned radial coordinate $\tilde{r}$ and radius $\tilde{R}$. We will also use the dimensionless axial coordinate $z=\tilde{z}/\tilde{L}$ in terms of the dimensioned axial coordinate $\tilde{z}$ and length $\tilde{L}$. Our dimensionless coordinates are then $r\in[0,1]$, $z\in[0,1]$, and an azimuthal angle $\phi\in[0,2\pi]$.  Our dimensionless radius $R=1$, which we will use as appropriate for clarity. The total free energy is the volume integral of the density, $F = \int_{V}f \mathrm{d}V$. 

The tensor $\lo = \boldsymbol{\underline{\underline{\delta}}} + (\zeta - 1)\boldsymbol{n_0}\otimes \boldsymbol{n_0}$ captures the initial orientations of molecules and cross-links in terms of the initial molecular director field $\no$, and the anisotropy parameter  $\zeta=\ell_\parallel/\ell_\perp$. $\boldsymbol{\underline{\underline{\delta}}}$ denotes the unit tensor, and $\otimes$ denotes a tensor product. $\zeta$ is the ratio between the projected length of cross-links in the directions parallel and perpendicular to $\no$. This gives
\begin{equation}
\lo = \begin{pmatrix} 1 & 0 & 0 \\ 0 & 1 + (\zeta-1) \sin^2\psi_0 & (1 - \zeta)\sin\psi\cos\psi_0 \\ 0 & (1 - \zeta)\sin\psi_0\cos\psi_0 & 1 + (\zeta-1) \cos^2\psi_0 \end{pmatrix}.
\end{equation}

For the post-strain director field, we again assume a double-twist structure:
$\n(r) = -\sin\psi(r) \boldsymbol{\hat{\phi}} + \cos\psi(r) \boldsymbol{\hat{z}}$.
The  post-strain structure of the elastomer is described by $\xlink = \boldsymbol{\underline{\underline{\delta}}} + (\zeta - 1)\boldsymbol{n}\otimes \boldsymbol{n}$, 
which leads to
\begin{equation}
\xlink^{-1} = \begin{pmatrix} 1 & 0 & 0 \\ 0 & 1 + (\zeta^{-1}-1) \sin^2\psi & (1 - \zeta^{-1})\sin\psi\cos\psi \\ 0 & (1 - \zeta^{-1})\sin\psi\cos\psi & 1 + (\zeta^{-1}-1) \cos^2\psi \end{pmatrix}.
\end{equation}

We consider an extension of the elastomer cylinder by a factor of $\lambda$, with extensional strain $\epsilon \equiv \lambda -1$, and impose incompressibility so that we have a radial compression by a factor of $\lambda^{-1/2}$. We also allow for a torsion deformation, where the two ends of the fibril are rotated around the $z$-axis with respect to each other by an angle $\tau L$.  Since the rotation along the cylinder will depend linearly on $z$, $\tau$ is therefore a torsion-gradient. We assume that $\lambda$ and $\tau$ are independent of position. Mathematically, we have
\begin{subequations}\label{deformation}
\begin{align}
z &\to\lambda z,\\
r &\to \lambda^{-1/2} r,\\
\phi &  \to \phi + \tau z.
\end{align}
\end{subequations}
From Fig.~\ref{fig:diagram}, we note that a positive $\tau$ will act to untwist (straighten) the molecular field lines for a positive angle function $\psi(r)$. We evaluate the deformation gradient tensor $\strain$ in Appendix~\ref{deformation_evaluation} using \cite{melnikov2018deformation}. The resulting tensor is
\begin{equation}\label{deformation_tensor}
\strain = \begin{pmatrix}
\lambda^{-1/2} & 0 & 0\\
0 & \lambda^{-1/2} & \tau \lambda^{-1/2} r\\
0 & 0 & \lambda
\end{pmatrix}.
\end{equation}
(This also agrees with the deformation gradient tensor used by Giudici and Biggins \cite{Giudici2021}.)
Using the tensors $\lo$, $\xlink$ and $\strain$, the free energy density $f$ is given by
\begin{widetext}
\begin{eqnarray}\label{eq:feq}
\frac{8 \zeta \lambda}{\mu} f &=& 1+ \lambda^3 + \zeta \big[6+\zeta+(2+\zeta) \lambda^3 \big]+r^2 (1+\zeta )^2  \tau ^2 
\\&&
+2 (-1+\zeta )   \sin[2 \psi_0] 
\left(-r (1+\zeta ) \tau -r (-1+\zeta ) \tau  \cos[2 \psi]-(-1+\zeta ) \lambda^{3/2} \sin[2 \psi]\right) \nn
\\&&
+\left(-1+\zeta ^2\right)  \left(\left(1-\lambda ^3+r^2 \tau ^2\right) \cos[2 \psi ]+2 r \lambda ^{3/2} \tau  \sin[2 \psi ]\right) \nn
\\&&
+\cos[2 \psi_0] \bigg[\left(-1+\zeta ^2\right) \left(-1+\lambda ^3+r^2 \tau ^2\right)+(-1+\zeta )^2 \left(-\left(1+\lambda ^3-r^2  \tau ^2\right) \cos[2 \psi]
+2 r \lambda ^{3/2} \tau  \sin[2 \psi]\right) \bigg] \nn.
\end{eqnarray}
\end{widetext}

We consider two different deformations: either an imposed torsion gradient $\tau$ corresponding to fibril ends that are ``clamped'' at a fixed rotation, or an equilibrium torsion gradient $\tau_\mathrm{eq}$ corresponding to fibril ends that can freely rotate. In both cases the extension factor $\lambda$ is fixed, but the post-strain twist angle function $\psi(r)$ is free to vary. In the first case, we must simply determine the post-strain twist angle function which minimizes the total free energy for a given $\tau$, by solving $\partial F/\partial \psi(r) = 0$. In the second case we must simultaneously solve for the $\psi(r)$ and $\tau_\mathrm{eq}$ which satisfy both $\partial F/\partial \psi(r) = 0$ and $\partial F/\partial \tau |_{\tau_{eq}}=0$. With azimuthal and translational symmetry, we have $F = 2 \pi \int_0^1 dr \, r f$.

\section{Equilibrium  with torsion}
\subsection{Self-consistent equations}
For a fixed torsion gradient $\tau$, we can locally minimize $f$ with respect to the post-strain twist-angle function $\psi(r)$. For $\zeta=1$, the elastomer is isotropic and $\partial f/\partial \tau$ directly determines $\tau=0$. For anisotropic elastomers, with $\zeta \neq 1$, setting $\partial f/\partial \psi = 0$ and solving for $\psi(r)$ yields Eq.~\eqref{psieq}, where the $r$-dependence of the function $\psi_0(r)$ is omitted for simplicity. When $\tau=0$, this reduces to the expression for $\psi(r)$ reported in \cite{Leighton2021chiral}. When $\tau$ is allowed to vary freely, the equilibrium value $\tau_\mathrm{eq}$ which locally minimizes the free energy is given by Eq.~\eqref{taueq}.
\begin{widetext}
\begin{equation}\label{psieq}
\psi(r) = \frac{1}{2}\cot^{-1}\left(\frac{(\zeta+1)\left[\lambda^3 - \tau^2 r^2 -1\right] + (\zeta-1)\left[\lambda^3 - \tau^2 r^2 +1\right]\cos(2\psi_0)+2(\zeta-1) \tau r\sin(2\psi_0)}{2\lambda^{3/2}\left[ (\zeta-1)\sin(2\psi_0) - (\zeta+1) \tau r - (\zeta-1) \tau r\cos(2\psi_0)\right]} \right),
\end{equation}
\begin{equation}\label{taueq}
\tau_\mathrm{eq} = \frac{\int_0^1\mathrm{d}r \, r^2 \left[ \left(\zeta^2-1\right)\left[\sin(2\psi_0) - \lambda^{3/2}\sin(2\psi)\right] + (\zeta-1)^2\left[\sin(2\psi_0)\cos(2\psi)- \lambda^{3/2}\cos(2\psi_0)\sin(2\psi)\right] \right]}
{\int_0^1\mathrm{d}r \, r^3 \left[ (\zeta+1)^2 + \left(\zeta^2-1\right)\left[\cos(2\psi_0) + \cos(2\psi)\right] + (\zeta-1)^2\cos(2\psi_0)\cos(2\psi) \right]}.
\end{equation}
\end{widetext}

In practice the total free energy $F$ must be used to distinguish local minima from maxima. To obtain good initial guesses for fast numerical calculation of extrema of $F$, we use an analytic approximation to the free energy~\eqref{eq:feq} for small $\psi$ and $\tau$ -- see Appendix~\ref{6th_order}. This 6th order approximation also allows us to quickly determine whether a given extrema corresponds to a local minimum or maximum of $F$.  We then numerically explore the full $F$ as a function of $\tau$ by computing $\psi$ using Eq.~\eqref{psieq} then computing $F$ by integrating Eq.~\eqref{eq:feq}. 

Previous modelling work indicates that the double-twist angle has non-trivial radial dependence that depends on the equilibrium vs non-equilibrium conditions of fibrillogenesis \cite{Cameron2020, Leighton2021non}. The above equations are valid for general $\psi(r)$, but for simplicity we restrict our attention to two cases that are often used as models of double-twist structure and which are reasonable approximations of more detailed models \cite{Cameron2020, Leighton2021non}. The first is a linear twist angle with radius, $\psi(r) = r \psi(R)$ -- where $r$ is measured in units of the fibril radius (shown in the rest of the Results). The second is a constant twist angle with radius, $\psi(r) = \psi(R)$ (shown in Appendix~\ref{constant_twist}).  Both the constant and linear twist models lead to qualitatively similar results, indicating a degree of robustness of our results to the details of $\psi(r)$. However, significant quantitative differences also indicate that detailed results will depend upon the specific $\psi(r)$ used. 

In order to make quantitative comparisons with biological collagen fibrils, we approximate tendon fibrils as having constant $\psi_0(r)$ with $\psi_0(R)\approx0.1$, and corneal fibrils as having linear $\psi_0(r)$ with $\psi_0(R)\approx 0.3$~\cite{Leighton2021non}. Based on previous work, we expect $\zeta\in[1.1,1.3]$~\cite{Leighton2021d}.

\subsection{Zero-strain limit} 
\begin{figure}[h!]
\includegraphics[width=\columnwidth]{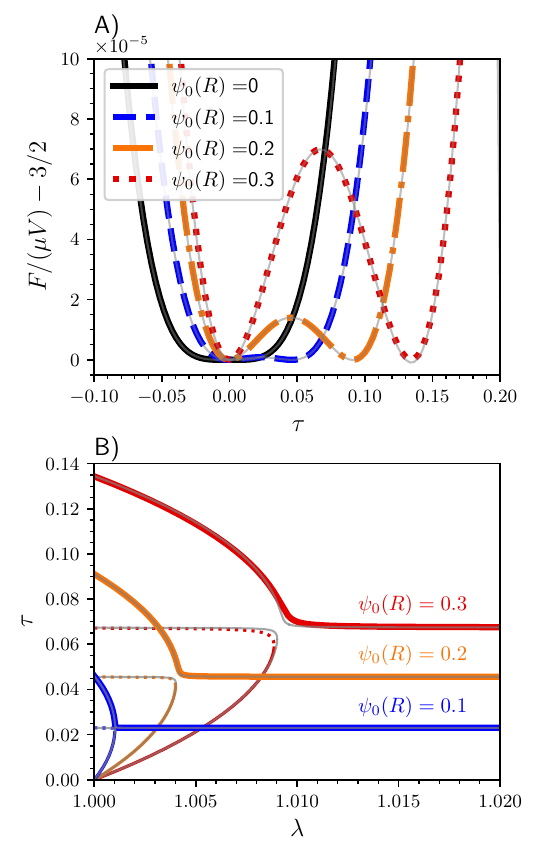}
\caption{\label{fig:F_Tau_Lin} A) Free energy $F$ in units of $\mu V$ vs. the torsion gradient $\tau$ at $\lambda=1$ (zero-strain) for  linear twist functions $\psi_0(r) = r\, \psi_0(R)$, with $\psi_0(R)$ indicated by the legend. B) Stable (solid lines, corresponding to local minima of $F$) and unstable (dotted lines, corresponding to local maxima of $F$) solutions for $\tau$ as a function of extension factor $\lambda$ for linear twist functions with $\psi_0(R)$ as indicated. The lower stable and unstable solutions merge as $\lambda$ increases, leaving only the upper stable (equilibrium) solution at high strains. Both: Coloured curves show solutions to the full equations \eqref{eq:feq} and \eqref{psieq}, while light grey curves show the 6th order approximation~\eqref{eq:6thorder}. We take $\zeta=1.3$.}
\end{figure}

Our anisotropic double-twist elastomer displays unexpectedly complex behaviour. We expect a vanishing torsion gradient $\tau$ for vanishing strains (with $\lambda \rightarrow 1$ and strain $\epsilon \equiv \lambda -1 \rightarrow 0$). However, at small strains we find multiple self-consistent solutions to Eqns.~\eqref{psieq} and \eqref{taueq}. To understand why, it is useful to examine the free energy $F$ as a function of $\tau$ -- as shown in Fig.~\ref{fig:F_Tau_Lin}A for $\lambda=1$. We observe a double-well form for the free energy, with the expected $\tau=0$ minimum but also a second local minimum at larger $\tau$, separated from the $\tau=0$ equilibrium by a free energy barrier. The accuracy of our 6th order expansion of $F$ in $\tau$ and $\psi$ (see Appendix~\ref{6th_order}) is illustrated by the grey lines in Fig.~\ref{fig:F_Tau_Lin}, where we see only small deviations from the full $F$ results even at $\psi_0(R)=0.3$. 

The non-zero torsion minimum of $F$ represents an alternative stable equilibrium. How large is the barrier between minima in units of $k_B T$? In Fig.~ \ref{fig:F_Tau_Lin}A, the free energy is in units of $\mu V$.  With Young's modulus $Y \approx 3 \mu$ (see below), we obtain a barrier height of approximately $25 k_B T$ for $\psi_0(R)=0.1$ with a cylinder of length $\tilde{L}=10 \mu m$ and radius $\tilde{R}=100nm$ and an elastic modulus $Y = 1 MPa$ typical of soft elastomers. For stiffer systems such as collagen \cite{Dutov2016} the barriers are much larger. Because of the large barriers, we expect that the mechanical history -- including the initial conditions -- of elastomeric systems would determine which stable branch would be observed.

\subsection{Bifurcation under strain} 
When the fibril is sufficiently strained, the values of $\tau$ corresponding to the two minima shown in Fig.~\ref{fig:F_Tau_Lin}A converge. Figure \ref{fig:F_Tau_Lin}B shows the stable (corresponding to local minima of $F$) and unstable (local maxima) values of $\tau$ as the fibril extension factor $\lambda$ is increased. The small-$\tau$ minimum and the local maximum collide and vanish at a critical  $\lambda_c$ in a saddle-node bifurcation, leaving only a large-$\tau$ solution as a global free energy minimum at larger $\lambda$. Despite their significant differences in free energy landscapes (see Appendix~\ref{constant_twist}) constant and linear twist fibrils exhibit qualitatively similar bifurcation behaviour of $\tau_\mathrm{eq}$ under extension.

\begin{figure}[t!]
\includegraphics[width=\columnwidth]{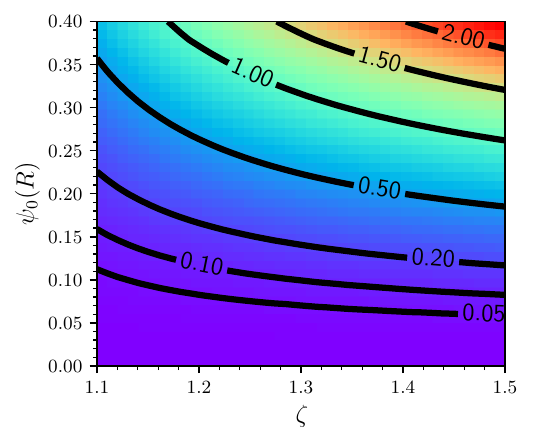}
\caption{\label{fig:CriticalStrain} The critical strain, $\epsilon_c \equiv \lambda_c-1$, in percent for a range of values of $\zeta$ and $\psi_0(R)$ for linear initial twist functions. For strains $\epsilon >\epsilon_c$, only a single high-torsion stable solution exists. For $\epsilon< \epsilon_c$ two stable and one unstable solutions exist, leading to a saddle-node bifurcation at $\epsilon_c$. 
}
\end{figure}

In Fig.~\ref{fig:CriticalStrain} we show the critical strain $\epsilon_c \equiv \lambda_c-1$ (in percent), above which only the large-$\tau$ equilibrium exists, for different values of $\psi_0(R)$ (assuming a linear twist function) and $\zeta$. As in Fig.~\ref{fig:F_Tau_Lin}B, increasing $\psi_0(R)$ leads to a higher critical strain. At higher $\psi_0(R)$ the cross-link anisotropy $\zeta$ significantly impacts the critical strain. For example, at $\psi_0(R)=0.3$, roughly corresponding to corneal fibrils \cite{Holmes2001}, varying $\zeta$ from $1.1$ to $1.5$ increases $\epsilon_c$ three-fold, suggesting that increased cross-link anisotropy leads to fibrils that are better protected from torsional instability.  Conversely, at smaller surface-twist $\psi_0(R)$ corresponding to tendon fibrils (with $\psi_0(R) \approx 0.1$) the bifurcation occurs at very small strain values ($\epsilon_c \lesssim 0.1 \%$) for all values of $\zeta$.

\subsection{Twist function}
At larger $\lambda$ (where $\epsilon>\epsilon_c$), where there is only one stable solution, we can simply expand Eqns.~\ref{psieq} and \ref{taueq} to first order in $\psi_0$, $\psi$, and $\tau$:
\begin{equation}\label{tausm}
\tau_\mathrm{sm} = 
\frac{4 (\zeta-1 ) }{ \zeta } 
\int_0^1 \tilde{r}^2   \Big[\psi_0(\tilde{r}) -\lambda ^{3/2}  \psi(\tilde{r})\Big] d\tilde{r},
\end{equation}
and
\begin{equation}
\psi_{sm} = \frac{\lambda^{3/2}}{\zeta \lambda^3 -  1}\Big[(\zeta - 1)\psi_0(r) - \zeta \tau r\Big], 
\label{psism}
\end{equation}
where `sm' indicates small-angle. 

\begin{figure}[h!]
\includegraphics[width=\columnwidth]{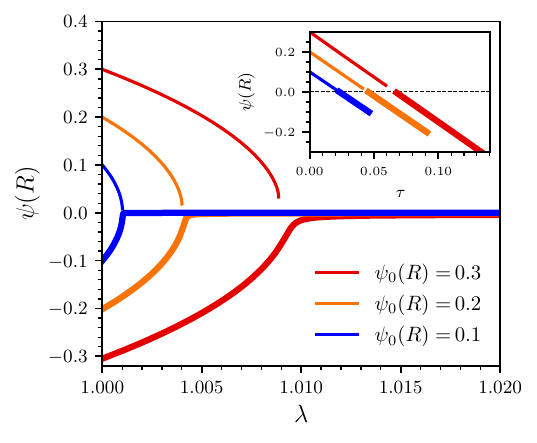}
\caption{\label{fig:psi_r}  Stable (outer branches) solutions for $\psi$ as a function of extension factor $\lambda$ for initial linear twist functions with $\psi_0(R)=0.1$, $\psi_0(R)=0.2$, and $\psi_0(R)=0.3$. Note that when the extensional strain exceeds $\epsilon_c$ the molecular twist is small ($\psi(r) \approx 0$). Inset: corresponding plots of $\psi(R)$ as a function of $\tau$ plotted parametrically by varying $\lambda$ along either stable branch. $\zeta =1.3$.}
\end{figure}
 
We then solve these linear equations to obtain self-consistent (`sc') solutions for $\psi(r)$ and $\tau$:
\begin{equation}\label{tausc}
\tau_{sc} =  \frac{4(\zeta-1)}{\zeta} \int_0^1 \tilde{r}^2\psi_0(\tilde{r})d\tilde{r},
\end{equation}
together with 
\begin{equation}\label{psisc}
\psi_{sc}(r) =  \frac{\lambda^{3/2} (\zeta-1)}{\zeta \lambda^3-1} \Big[ \psi_0(r) - 4r\int_0^1 \tilde{r}^2\psi_0(\tilde{r})d\tilde{r} \Big].
\end{equation}
We see that $\psi_{sc}(r)$ is linear in $r$ for linear or constant $\psi_0(r)$. We also have that $\int \tilde{r}^2 \psi_{sc}(\tilde{r}) d\tilde{r} = 0$. As a result, for linear twist with $\psi_0(r)=r \psi_0(R)$, then $\psi_{sc}(r)=0$ -- i.e. any initial linear twist is exactly unwound at higher strain.  For constant twist, with $\psi_0(r) = \psi_0(R)$, then $\psi \propto \psi_0(R)(1-4r/3)$ -- so that surface twist has the opposite sign from $\psi_0$ for $\epsilon>\epsilon_c$! 

More generally, Fig.~\ref{fig:psi_r} indicates how the surface twist  $\psi(R)$ varies under strain; the surface twist  exhibits the same bifurcation behaviour shown in Fig.~\ref{fig:F_Tau_Lin}B. The symmetry about $\psi(R)=0$ is broken by the choice of initial twist function $\psi_0$. The inset indicates that $\psi(R)$  decreases approximately linearly with torsion $\tau$, echoing the small $\tau$ limit of Eqn.~\ref{psism}. 

\subsection{Mechanical properties} 

\begin{figure}[h!]
\includegraphics[width=\columnwidth]{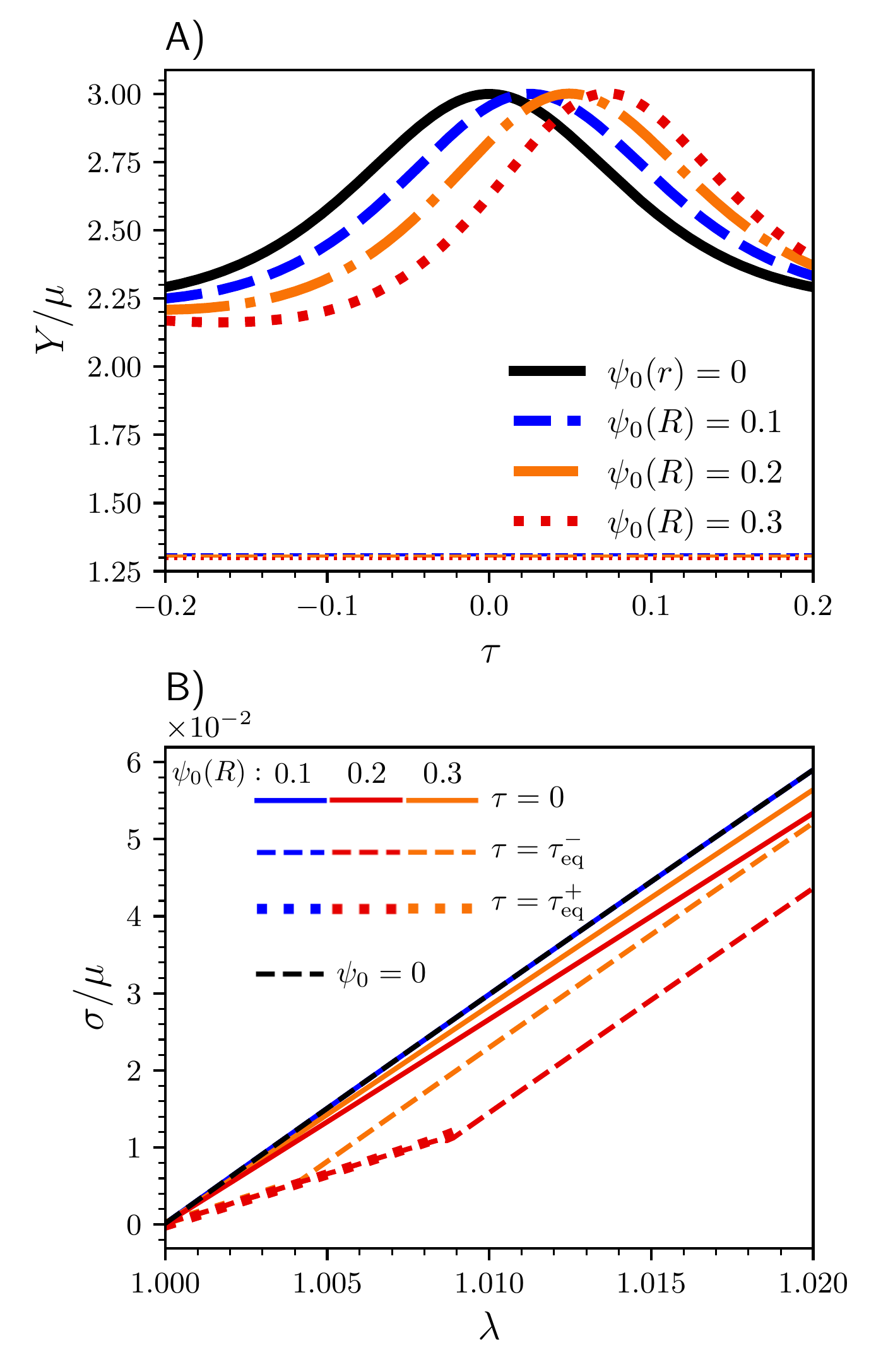}
\caption{\label{fig:Mechanics} A) The Young's Modulus $Y/\mu = \partial^2(F/\mu V)/\partial \lambda^2$ at $\lambda=1$ as a function of clamped $\tau$ for different linear twist angle functions. Horizontal lines show the Young's modulus when boundaries are unclamped and thus $\tau=\tau_\mathrm{eq}$ on the lower and upper branches (identical, thin lines). B) Stress ($\sigma = \partial (F/\mu V)/\partial \lambda$) vs strain ($\lambda=1+\epsilon$) curves for either clamped ends ($\tau=0$), or unclamped ends ($\tau=\tau_\mathrm{eq}$), indicated by thin dashed lines for the lower branch ($\tau=\tau_\mathrm{eq}^-$) and thick dotted lines for the upper branch ($\tau=\tau_\mathrm{eq}^+$) -- i.e.  for $\lambda < \lambda_c$.). The twist function $\psi_0(r)$ is linear. $\zeta=1.3$. }
\end{figure}

Torsion significantly changes the mechanical properties of elastomers. Consider fibrils with clamped ends such that a fixed initial torsion gradient $\tau$ is maintained during an extension. Figure \ref{fig:Mechanics}A shows the resulting Young's modulus measured when such a fibril is strained -- where $Y/\mu = \partial^2(F/\mu V)/\partial \lambda^2$ at $\lambda=1$. For both constant and linear twist angle functions, pre-torsion of the fibril can significantly alter the Young's modulus. For example, for a constant twist of $0.1\mathrm{rad}$, roughly corresponding to tendon fibrils, varying the imposed torsion gradient $\tau$ from $-0.1$ to $0.1$ results in an almost $50\%$ difference in measured Young's modulus. 

The effects of allowing ends to freely rotate are even more dramatic. The thin horizontal lines in Fig.~\ref{fig:Mechanics}A indicate the Young's modulus of a fibril with unclamped boundary conditions (with $\tau=0$ at $\lambda=1$, following the upper or lower stable branch in Fig.~\ref{fig:F_Tau_Lin}B). Interestingly, the Young's modulus are the same for the two stable branches. The Young's modulus of unclamped fibrils falls far below the Young's modulus of fibrils with clamped ends, by a factor of $2-3$ depending on the initial torsion gradient and initial twist angle function. 

The bifurcation at $\lambda_c$ also affects the mechanical properties. This is illustrated in Fig.~\ref{fig:Mechanics}B, which shows the stress-strain curves of fibrils with freely rotating boundary conditions ($\tau = \tau_\mathrm{eq}$) with dashed lines. Fibrils clamped with $\tau=0$ are shown for reference with solid lines.  At $\lambda = \lambda_c$ we see an abrupt stiffening of the fibril (note that the Young's modulus is the slope  $Y/\mu = \partial \sigma /\partial \lambda$). While the torsion dependence of mechanical properties is also seen with constant twist fibrils, the details differ significantly (see Fig.~\ref{fig:Mechanics_Const} in Appendix~\ref{constant_twist}). 

\section{Discussion}
We have explored the effects of free torsion on the structural and mechanical properties of strained double twist elastomeric cylinders. Allowing for freely-rotating ends leads to qualitatively different behavior from fixed ends. For small extensional strains there are two stable torsion values. At a critical strain the system undergoes a bifurcation, where the torsion and twist angle both change discontinuously.  Above the critical strain only a single stable torsion value is observed. 

Torsion, twist angles, and mechanical properties all change discontinuously at the critical strain, but they also continuously depend on strain -- particularly below the critical strain. To provide some perspective on the expected scale of physical effects, we will focus our attention on collagen fibrils.

The torsion gradients $\tau$ appear small, but can correspond to large relative twist angles between fibril ends since we measure length in radial units so $\Delta \phi = \tau \tilde{L}/\tilde{R}$. For a tendon collagen fibril of radius $\tilde{R}=100nm$ and length $\tilde{L}=10 \mu m$ we have $\tilde{L}/\tilde{R} \approx 100$. A torsion gradient of $\tau=0.02$ (corresponding to a tendon fibril with linear twist and $\psi(R)=0.1$ under a $0.1\%$ strain) then leads to an end-to-end rotation of $\gtrsim 100^\circ$!  

Freely-rotating ends also lead to different mechanical properties. As shown in Fig.~\ref{fig:Mechanics}A, fibrils with freely-rotating ends can have $2-3$ times smaller Young's Modulus than fibrils with clamped ends.  The stress-strain curves of fibrils with freely-rotating ends (Fig.~\ref{fig:Mechanics}B) also shows much lower stresses at small strains, a phenomenon similar to the ``toe region" seen in recent single-fibril studies ~\cite{nalbach2022instrument}. 

Such dramatic physical effects would provide a window into the internal structure of collagen fibrils, particularly the magnitude ($\psi_0(R)$, see Figs.~\ref{fig:F_Tau_Lin}-\ref{fig:Mechanics}) and functional form ($\psi_0(r)$, compare with Figs.~\ref{fig:F_Tau_Const}-\ref{fig:Mechanics_Const}) of the double-twist function. Free-rotation of single fibrils \emph{in vitro} may be possible with magnetic-bead mechanical coupling \cite{nalbach2022instrument}. Our freely-rotating equilibrium results should apply if \emph{either} fibril end is freely-rotating.

While we are not aware of existing experimental studies that have allowed free rotation of collagen fibrils, the Young's modulus is very sensitive to the magnitude and sign of any pre-existing torsion gradient. This may explain some of the variability of Young's modulus reported for tensile tests of single fibrils \emph{in vitro} where fibrils are glued in the dry state and rehydrated before testing  \cite{Svensson2013, Svensson2021}. If the degree of torsion is uncontrolled and variable due to the preparation process, this could lead to a $25\%$ variability of the measured Young's modulus (see Fig.~\ref{fig:Mechanics}). 

Bundles of fibrils both \emph{ex vivo} or \emph{in vivo} will have extensive interfibrillar bonds and cross-links that will inhibit free fibril rotation. Any fibril-level torque would be aggregated via interfibrillar linkages \cite{Watanabe2016} and lead to shape effects at higher hierarchical levels.  We expect torque from inhibited fibril torsion to couple to fibril shape \cite{Kalson2015, Bozec2007, Safa2019} and also to higher hierarchical structures such as fibers \cite{Kalson2011, Thorpe2013, Buchanan2017}, wave-like structures \cite{Vidal2003}, or crimps \cite{Franchi2010}. Indeed, torsion effects of fascicles are observed under tension. As an example, 10mm samples from mouse tendon fascicles rotated an average of $51^\circ$ under $1\%$ strain \cite{Buchanan2017}. 

Torque-shape effects could vary between different developmental stages. Block-face electron microscopy reconstruction of mouse tail tendon sections \cite{Kalson2015} shows changing helical shapes of single fibrils during development. In the embryo (day 15), individual fibrils have no clear helical pitch. At birth, each fibril follows a helical path with a contour length of 17 microns per turn with radius $\tilde{R} \approx 23 nm$. After 6 weeks,  fibrils follow an helical path with a contour length of 100 microns per turn with radius $\tilde{R} \approx 80nm$. Under the simplifying assumption that the shape distortion is comparable to the frustrated torsion, this corresponds to $\tau$ increasing from 0 (embryonic) to 0.009 at birth and then decreasing to 0.005 after 6 weeks. Mature tendon fibrils have an estimated $\psi_0(R) \approx 0.1$ \cite{Reale1981}, so these results are consistent with fibrils being strained by less than $0.1\%$ at birth.  Our model then predicts that the molecular tilt at the surface $\psi_0(R)$ will be approximately linear in strain -- so half the value at birth compared to after 6 weeks (see Fig.~\ref{fig:psi_r} inset). 

While we are inclined to assume that fibrils are formed with zero torsion $\tau$,  transient strains experienced during development could drive fibrils through the bifurcation to high-torsion behavior -- though presumably not once interfibrillar connections are made that would fix $\tau$. For the small double-twists of tendon fibrils with $\psi_0(R) \approx 0.1$, we would expect a very small critical strain $\epsilon_c$ (see Fig.~\ref{fig:CriticalStrain}) -- so both stable and metastable bifurcations are plausible. The timing of applied strain during development is therefore important. For the larger double-twists of corneal fibrils with $\psi_0(R) \approx 0.3$ it seems possible that the critical bifurcation could be avoided during development. 

Experimental studies are needed to explore torsion effects for both single fibrils and for assemblies of fibrils at the tissue level. The relationship between torsion, torque, and shape effects needs to be better explored for collagen -- as it can be complex \cite{Michaels2020, Love}. Single-fibril experiments with defined non-zero torsion should show strong mechanical effects even without any double-twist (see Fig.~\ref{fig:Mechanics}A). Allowing freely-rotating ends, perhaps with magnetic bead attachments \cite{nalbach2022instrument}, would lead to more dramatic effects. Fibrils with larger twist angles $\psi_0(R)$ should show large torsion effects (or torsion-induced shape effects) that exhibit hysteresis around the critical strain. Tissue experiments may be easier to accomplish (see e.g. \cite{Buchanan2017}), but may be harder to interpret if the underlying fibrils have a distribution of initial torsions.  Isolated \emph{in vitro} assembled  fibrils with a defined mechanical history would provide the most direct test of our results.

The recent theoretical study of Giudici and Biggins \cite{Giudici2021} explored the torsion/extension coupling of fibers through a nematic/isotropic phase-transition. Our work complements theirs, since we work entirely within a liquid-crystalline ordered phase and explore mechanical effects. Using finite-element methods (FEM), they demonstrated that frustrated torsion can have a strong effect on fiber shape. Applying similar FEM in our context would provide greater insight into the coupling of single fibril torque into shape effects of both fibrils and tissue. 

Finally, in this work we have only considered extensional strains. Previous work has found rich behaviour under compression for fibrils with clamped ends ($\tau=0$) -- both experimentally \cite{peacock2020buckling} and theoretically \cite{Leighton2021chiral}. We expect that compressed fibrils with free torsion could exhibit even more complex behavior.

\emph{Acknowledgments}.---We thank the Natural Sciences and Engineering Research Council of Canada (NSERC) for operating Grants RGPIN-2018-03781 (LK) and RGPIN-2019-05888 (ADR). MPL thanks NSERC for summer fellowship support (USRA-552365-2020), and CGS Masters and Doctoral fellowships. 

\bibliography{main.bib}
\bibliographystyle{rsc} 

\appendix
\section{The deformation gradient tensor}\label{deformation_evaluation}

The most general deformation possible without breaking azimuthal symmetry is given by 
\begin{subequations}
\begin{equation}
z\to\alpha(r,z),
\end{equation}
\begin{equation}
r\to\gamma(r,z),\text{ and}
\end{equation}
\begin{equation}
\phi \to \phi + \Delta \phi(r,z).
\end{equation}
\end{subequations}
This allows for e.g. inhomogeneities in $z$ or $r$. We use this deformation to derive the general deformation gradient tensor $\strain$ in a cylindrical geometry. 

Following Ref.~\cite{melnikov2018deformation}, in curvilinear coordinates the deformation gradient tensor can be written as
\begin{equation}\label{grad_v}
\begin{aligned}
\strain  & = \nabla \boldsymbol{v}\\
& = \sum_{j,k,m,d=1}^3 \left( \frac{\partial v^k}{\partial X^j} + v^d\Gamma_{dm}^k\frac{\partial x^m}{\partial X^j}\right)\boldsymbol{g}_k(\boldsymbol{x})\otimes\boldsymbol{G}^j(\boldsymbol{X}),
\end{aligned}
\end{equation}
where the vectors $\bm{X}$ and $\bm{v}$ describe the position vector in the original and deformed states, respectively:
\begin{subequations}
\begin{align}
\bm{X} & = r \bm{e}_r + z\bm{e}_z\\
\bm{v} & = \gamma(r,z) \bm{e}_r + \alpha(r,z)\bm{e}_z.
\end{align}
\end{subequations}
We also have
\begin{subequations}
\begin{align}
\bm{G}(\bm{X}) & = \bm{e}_r + \bm{e}_\phi/r + \bm{e}_z\\
\bm{g}(\bm{x}) & = \bm{e}_r + \gamma(r,z) \bm{e}_\phi + \bm{e}_z.
\end{align}
\end{subequations}
The indices $1$, $2$, and $3$ denote the $r$, $\phi$ and $z$ coordinates respectively, and $\Gamma_{dm}^k$ are the Christoffel symbols for cylindrical geometry, only three of which are non-zero:
\begin{subequations}
\begin{align}
\Gamma_{12}^2 & = \Gamma_{21}^2 = 1/r,\\
\Gamma_{22}^1 & = -r.
\end{align}
\end{subequations}

The components $X^j$, $x^m$, and $v^k$ in equation \ref{grad_v} are
\begin{subequations}
\begin{align}
X^1 & = r,\\
X^2 & = \phi,\\
X^3 & = z,\\
x^1 & = \gamma(r,z),\\
x^2 & = \phi + \Delta \phi(r,z),\\
x^3 & = \alpha(r,z),\\
v^1 & = \gamma(r,z),\\
v^2 & = 0,\\
v^3 & = \alpha(r,z).
\end{align}
\end{subequations}

We can now proceed to evaluate the deformation gradient tensor. This is tedious but straightforward, yielding
\begin{equation}
\strain = \begin{pmatrix}
\gamma_r & 0 & \gamma_z\\
\gamma\Delta \phi_r & \gamma/r & \gamma\Delta \phi_z\\
\alpha_r & 0 & \alpha_z
\end{pmatrix},
\end{equation}
where the subscripts indicate partial derivatives, e.g. $\gamma_r \equiv \partial \gamma/ \partial r$.

Taking the deformation specified in the text (Eq.~\eqref{deformation}), we have $\gamma(r,z)=r/\lambda^{1/2}$ and so $\gamma_r = \lambda^{-1/2}$, $\Delta \phi(r,z) = \tau  z$ and so $\Delta \phi_z = \tau$, and $\alpha(r,z) = \lambda z$ and so $\alpha_z = \lambda$ -- with $\gamma_z=\Delta \phi_r=\alpha_r =0$. This then leads to  Eqn.~\ref{deformation_tensor}. This final result is consistent with that obtained by Ogden~\cite[page 113]{Ogden1997}, and was also used by Giudici and Biggins \cite{Giudici2021}.

\section{6th order approximation} \label{6th_order}
In the text and figures we made use of a 6th order approximation for the free energy. This is computed by expanding the free energy density~\eqref{eq:feq} to 6th order simultaneously in $\psi_0$, $\psi$, and $\tau$ (holding these three quantities proportional to each other). This procedure yields an unwieldy but analytical expression for the free energy density, which is used to compute thin grey lines in Fig.~\ref{fig:F_Tau_Lin}, and provides an initial guess to speed up numerical computation of the full free energy landscape.

The 6th order expansion for linear twist functions $\psi_0(r)=\psi_0^R r$ is 
\begin{equation}\label{eq:6thorder}
F =  \frac{a_1\tau +\frac{a_2}{2}\tau^2 +\frac{a_3}{3}\tau^3 +\frac{a_4}{4}\tau^4 +\frac{a_5}{5}\tau^5 +\frac{a_6}{6}\tau^6}{3 \left(\zeta \lambda^3-1\right)^3}.
\end{equation}
The coefficients are 
\footnotesize
\begin{widetext}
\begin{subequations}
\begin{align*}
a_1 & = \frac{1}{60 \lambda} (\zeta -1) \left(\lambda ^3-1\right) \psi_0^R \Biggl(-90 \left(\zeta 
   \lambda ^3-1\right)^2
    +40 (\psi_0^R)^2 \left(\lambda ^3
   \left(\zeta  \left(\zeta  \left(\lambda ^3+3\right)-8\right)+3\right)+1\right)  \\ \nonumber
   & 
   -\frac{3 (\psi_0^R)^4 \left(2 \zeta ^4 \lambda ^{12}+\zeta 
   (\zeta  ((67-15 \zeta ) \zeta -105)+45) \lambda ^9+3 (5 (\zeta -2) \zeta +3) (\zeta 
   (3 \zeta -10)+5) \lambda ^6+(\zeta  (15 \zeta  (3 \zeta -7)+67)-15) \lambda
   ^3+2\right)}{\left(\zeta  \lambda ^3-1\right)^2}
   \Biggr),\\
a_2 & = \frac{1}{4 \lambda}  \Biggl(6 \zeta  \left(\lambda ^3-1\right) \left(\zeta  \lambda
   ^3-1\right)^2    -4 (\zeta -1) (\psi_0^R)^2
   \left(\zeta ^2 \lambda ^9+(\zeta -4) \zeta  \lambda ^6-3 (2 (\zeta -2) \zeta +1)
   \lambda ^3-1\right)\\ 
   &+\frac{(\zeta -1) (\psi_0^R)^4 \left(\lambda ^3 \left(\zeta 
   \left(\zeta ^3 \lambda ^{12}-\zeta  \left(2 \zeta ^2+3\right) \lambda ^9-45 \zeta
   ^2+(\zeta  (\zeta  (30 \zeta -119)+156)-57) \lambda ^6+(188-3 \zeta  (\zeta  (15 \zeta
   -76)+117)) \lambda ^3+102 \zeta -67\right)-30 \lambda
   ^3+15\right)-1\right)}{\left(\zeta  \lambda ^3-1\right)^2}
\Biggr),\\
a_3 & = \frac{3}{2} (\zeta -1)^2 \lambda ^2 \psi_0^R \left(\frac{(\psi_0^R)^2 \left(\zeta 
   \left(2 (3-2 \zeta ) \zeta  \lambda ^6+(5 \zeta  (3 \zeta -8)+21) \lambda ^3+15 \zeta
   -16\right)+3\right)}{\left(\zeta  \lambda ^3-1\right)^2}-4 \zeta \right),\\
a_4 & = \frac{1}{2} (\zeta -1) \zeta  \lambda ^2 \left(4 \zeta +\frac{3 (\zeta -1) (\psi_0^R)^2
   \left(\zeta  \left(\lambda ^3 \left(\zeta  \left(\lambda
   ^3-15\right)+19\right)-15\right)+10\right)}{\left(\zeta  \lambda ^3-1\right)^2}\right),\\
a_5 & = \frac{45 (\zeta -1)^2 \zeta ^2 \lambda ^2 \psi_0^R \left(\zeta  \lambda
   ^3+1\right)}{4 \left(\zeta  \lambda ^3-1\right)^2},\\
a_6 & = -\frac{9 (\zeta -1) \zeta ^3 \lambda ^2 \left(\zeta  \lambda ^3+1\right)}{4 \left(\zeta 
   \lambda ^3-1\right)^2}.
\end{align*}
\end{subequations}
\end{widetext}
\normalsize
In a manner analogous to that described above, we also derive a separate small-angle expansion for constant twist angle functions $\psi_0(r) = \psi_0^R$, which is then used to compute the thin grey lines in Fig.~\ref{fig:F_Tau_Const}. This expansion (not shown) is similarly unwieldy but easily derivable from Eq.~\eqref{eq:feq}.

\section{Constant twist results $\psi_0(r) = \psi_0(R)$} \label{constant_twist}
While in the main text we focused on primarily on linear initial twist angle functions, constant twist angle functions are also of significant interest, being thought to roughly correspond to the biologically-relevant example of tendon collagen fibrils. 

\begin{figure}[h]
\includegraphics[width=\columnwidth]{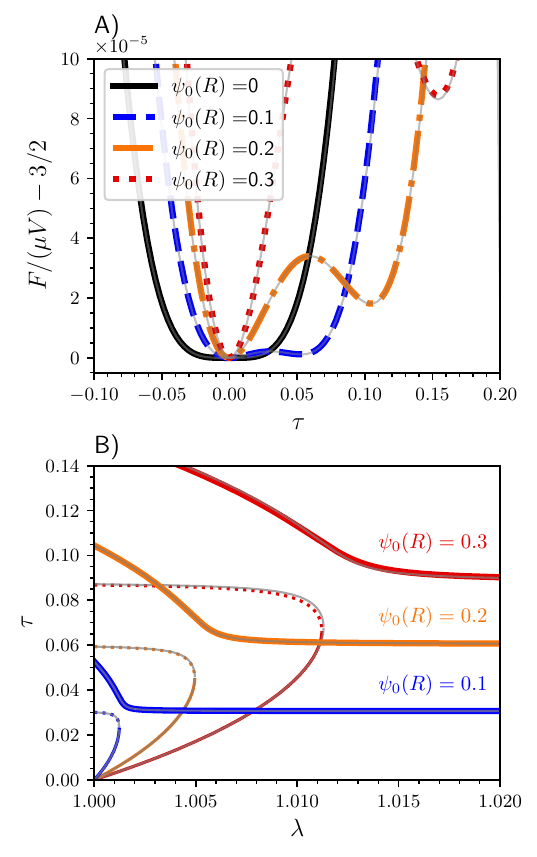}
\caption{\label{fig:F_Tau_Const} A) Free energy $F$ as a function of the torsion gradient $\tau$ at $\lambda=1$ (zero-strain) for constant twist functions $\psi_0(r) = \psi_0(R)$. B) Stable (solid lines, corresponding to local minima of $F$) and unstable (dotted lines, corresponding to local maxima of $F$) solutions for $\tau$ as a function of extension factor $\lambda$ for several different constant initial twist functions. The lower stable and unstable solutions merge as $\lambda$ increases, leaving only the upper stable solution at high strains. Both: Coloured curves show solutions to the full equations \eqref{eq:feq} and \eqref{psieq}, while light grey curves show the 6th order approximation. $\zeta=1.3$.}
\end{figure}

Figure \ref{fig:F_Tau_Const}A shows the free energy landscape as a function of $\tau$ at zero strain for constant twist angle functions. While the free energy landscape has a double-well shape as in the linear twist case (Fig.~\ref{fig:F_Tau_Lin}), the two minima are highly asymmetric for constant twist functions. Figure \ref{fig:F_Tau_Const}B then shows the same three-branched bifurcation of $\tau$ as strain is increased; this is qualitatively the same as the behaviour seen for linear twist in Fig.~\ref{fig:F_Tau_Lin}B.

\begin{figure}[h]
\includegraphics[width=\columnwidth]{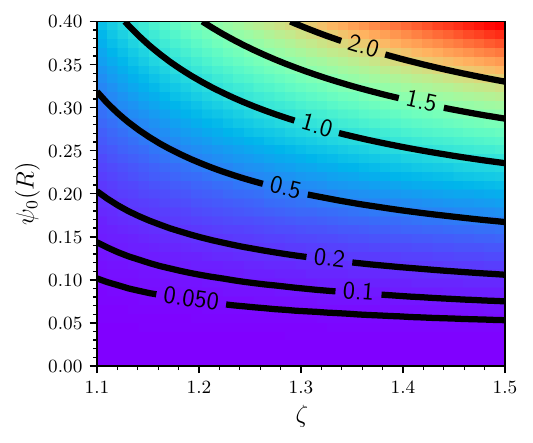}
\caption{\label{fig:CriticalStrain_Const} The critical strain, $\epsilon_c=\lambda_c-1$, in percent for a range of values of $\zeta$ and $\psi_0(R)$ for constant initial twist functions. When the extensional strain exceeds $\epsilon_c$ the molecular twist is small with $\psi(r) \approx 0$.}
\end{figure}

Figure \ref{fig:CriticalStrain_Const} shows the critical strain as a function of initial surface twist $\psi_0(R)$ and anisotropy $\zeta$ for constant twist. We see no qualitative differences with respect to the linear case (Fig.~\ref{fig:CriticalStrain}), and only small quantitative differences.

\begin{figure}[h]
\includegraphics[width=\columnwidth]{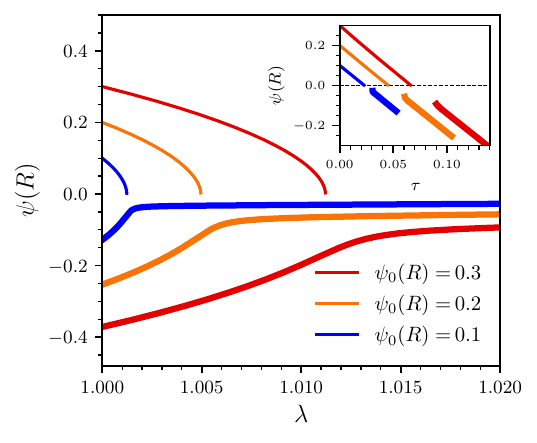}
\caption{\label{fig:psi_r_Const}  Stable (outer branches) solutions for $\psi$ as a function of extension factor $\lambda$ for initial constant twist functions with $\psi_0(R)=0.1$, $\psi_0(R)=0.2$, and $\psi_0(R)=0.3$. Dark thick curves show $\psi(R)$, while lighter, thinner curves show $\langle \psi(r)\rangle$. Inset: $\psi(R)$ as a function of $\tau$, plotted parametrically for different values of $\lambda$. $\zeta=1.3$.}
\end{figure}

Figure \ref{fig:psi_r_Const} shows the strained twist angle $\psi$ as a function of strain for constant twist. While the behaviour is qualitatively similar to that seen for linear twist in the main text (Fig.~\ref{fig:psi_r}), the bifurcation for constant twist involves a much larger jump in $\psi$ when a fibril on the upper branch reaches the critical strain.

\begin{figure}[h]
\includegraphics[width=\columnwidth]{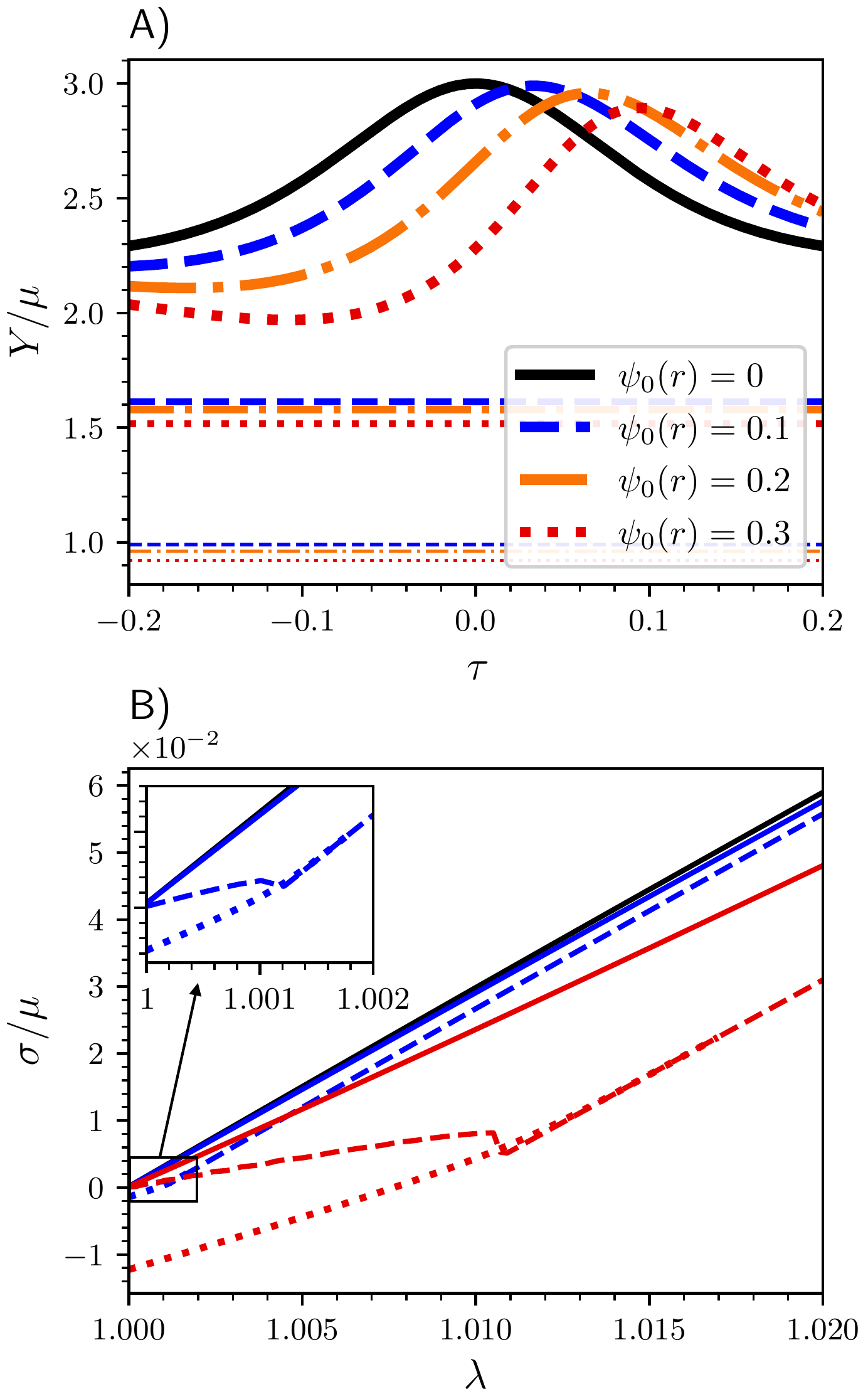}
\caption{\label{fig:Mechanics_Const} A) The Young's Modulus $Y/\mu = \partial^2(F/\mu V)/\partial \lambda^2$ as a function of clamped $\tau$ for different constant twist angle functions. Horizontal lines show the Young's modulus when boundaries are unclamped and thus $\tau=\tau_\mathrm{eq}$ on the lower (thin) and upper (thick) branches. B) Stress ($\sigma = \partial (F/\mu V)/\partial \lambda$) vs strain ($\lambda=1+\epsilon$) curves for either clamped ends ($\tau=0$, indicated by solid lines) or freely rotating ends ($\tau = \tau_\mathrm{eq}$, indicated by dashed lines for the lower branch and dotted lines for the upper branch.). The twist function $\psi_0(r)$ is constant with either $\psi_0(R)=0.1$ (blue curves) or $\psi_0(R)=0.3$ (red curves). We take $\zeta=1.3$. Inset: a close-up view of the stress-strain behaviour for $\psi_0(R)=0.1$ near the critical strain (red curves omitted for clarity).}
\end{figure}

Finally, Fig.~\ref{fig:Mechanics_Const} shows the mechanical properties for constant twist. Unlike the case of linear twist (Fig.~\ref{fig:Mechanics}), for constant twist we see significant differences between the mechanical properties of the upper and lower stable branches. While the Young's modulus of the lower branch is still $\sim50\%$ lower for the lower branch with unclamped boundaries than for clamped boundaries, Fig.~\ref{fig:Mechanics_Const}A shows that the upper branch lies in between, roughly $\sim50\%$ stiffer than the lower branch. Figure \ref{fig:Mechanics_Const}B shows even starker differences for the stress-strain curves. While the lower branch at constant twist (dashed lines) exhibits qualitatively the same change change in Young's modulus at the critical strain as seen for linear twist, the  Young's modulus of the upper branch (dotted lines) remains roughly constant across the critical strain.

\clearpage
\end{document}